\documentclass[aps,pre,
onecolumn,
twocolumn,
10pt,
notitlepage,
showpacs,floatfix,nofootinbib,
superscriptaddress
]{revtex4-2}
\usepackage{amsmath}
\usepackage{amsfonts}
\usepackage{amssymb}
\usepackage{graphicx}
\usepackage{dcolumn}
\usepackage{bm}
\usepackage{color}
\usepackage{ulem}
\usepackage{float}
\usepackage{orcidlink}
\usepackage{hyperref}
\hypersetup{colorlinks=true,
	linkcolor=blue,
	urlcolor=blue,
	citecolor=blue,
	pdfhighlight=/N
}

\newcommand{\kmax}{k_\text{max}}

\newcommand{\kc}{k_\text{c}}

\newcommand{\av}[1]{\left\langle #1 \right\rangle}

\begin{document}

\title{Epidemic spreading on multigraphs}

\author{Paulo H. Lorenzoni\orcidlink{0009-0006-1698-355X}}
\affiliation {Departamento de Física, Universidade Federal de Viçosa, 36570-900 Viçosa, Minas Gerais, Brazil}
\author{Wesley Cota\orcidlink{0000-0002-8582-1531}}
\affiliation {Departamento de Física, Universidade Federal de Viçosa, 36570-900 Viçosa, Minas Gerais, Brazil}
\author{Francisco A. Rodrigues\orcidlink{0000-0002-0145-5571}}
\affiliation{Instituto de Ci\^{e}ncias Matem\'{a}ticas e de Computa\c{c}\~{a}o, Universidade de S\~{a}o Paulo, S\~{a}o Carlos, SP 13566-590, Brazil}
\author{{Silvio C. Ferreira\orcidlink{0000-0001-7159-2769}}}
\affiliation {Departamento de Física, Universidade Federal de Viçosa, 36570-900 Viçosa, Minas Gerais, Brazil}
\affiliation{Instituto de Ci\^{e}ncias Matem\'{a}ticas e de Computa\c{c}\~{a}o, Universidade de S\~{a}o Paulo, S\~{a}o Carlos, SP 13566-590, Brazil}

\begin{abstract}  

\noindent 
Multigraphs are graphs in which multiple links between pairs of nodes are allowed, whereas they are forbidden in simple graphs, the latter being widely used in network science. Simple graphs generated by the configuration model have served as a benchmark for validating theoretical approaches to dynamical processes on networks. However, generating large scale-free networks with degree exponent $\gamma<3$ introduces uncontrolled disassortative correlations and severe computational limitations due to the prohibition of reconnecting hubs. These constraints do not exist in multigraphs. We investigate how multiple connections affect epidemic spreading by comparing several epidemic models exhibiting an active steady state on simple graphs and multigraphs sharing the same degree sequence and natural upper cutoff. By analyzing epidemic thresholds, finite-size scaling, and localization, we show that differences between simple graphs and multigraphs emerge only when epidemic activity can persist on isolated hubs (star subgraphs) for times exponentially long in the hub degree. Our results remove a methodological barrier to the study of dynamical processes on large scale-free networks.
\end{abstract}


\maketitle 

\textit{Introduction}. Complex systems are almost universally represented, at some level, by networked structures~\cite{barabasi2016network}, where components are represented by nodes and their interactions by the edges of a graph. While modern network science has progressively shifted its focus toward more sophisticated representations, including hypergraphs~\cite{Boccaletti2023}, temporal networks~\cite{Masuda2016}, and multilayer structures~\cite{Aleta2026}, standard static pairwise graphs are still valuable tools for understanding dynamical processes on networked systems. The list of relevant dynamical processes on networks is extensive, including epidemic models~\cite{PastorSatorras2015}, rumor propagation~\cite{Moreno2004,FerrazdeArruda2022}, social dynamics~\cite{Starnini2025}, synchronization~\cite{Rodrigues2016}, among many others~\cite{Dorogovtsev2022,Barrat2008}. High heterogeneity and structural diversity are intrinsic to most real networks~\cite{barabasi2016network}, and their role in dynamical processes has been extensively investigated using mean-field approximations~\cite{Wang2017,Starnini2025}. The validation and accuracy of these theoretical approaches require numerical simulations on synthetic networks with controlled structural properties~\cite{PastorSatorras2015,FerrazdeArruda2022,Silva2019}. Indeed, even simple dynamical models on networks, such as the Harris contact process~\cite{Harris1974}, have led to intense debate regarding the interplay between dynamics and structure~\cite{Hong2007,Castellano2006,Ha2007,Castellano2007,Castellano2008,Ferreira2011}, where simulations play a central role. More remarkably, the susceptible-infected-susceptible (SIS) epidemic model has been the subject of two decades of intensive investigation to elucidate the nature of its activation mechanisms~\cite{Goltsev2012,Boguna2013,Lee2013,Castellano2020,Mata2015}, especially in the non-scale-free regime with degree exponent $\gamma > 3$.

\begin{figure}[hbt]
	\centering
	\includegraphics[width=1\linewidth]{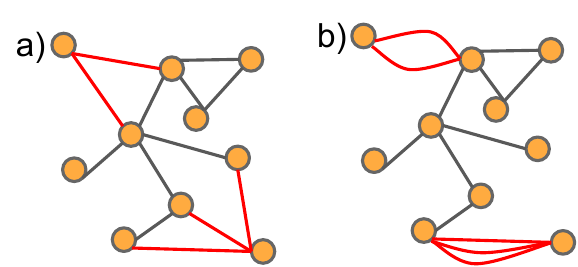}
	\caption{Examples of (a) simple graph and (b) multigraph. Nodes are represented by circles and edges by lines connecting them. Red links were transformed into multiedges. }
	\label{fig:multigraph}
\end{figure}

Since the pioneering simulation studies of dynamical processes on synthetic complex networks~\cite{Pastor-Satorras2001,Moreno2004,Holme2002}, it has become customary, mainly for theoretical convenience, to disregard multiple connections, despite their presence in real-world networks. Networks without multiple edges (multiedges) are called \textit{simple graphs}, whereas \textit{multigraphs}~\cite{Bollobas1998} refer to networks in which multiple connections are allowed; see Fig.~\ref{fig:multigraph}. A widely used model for generating random networks with a prescribed degree distribution is the Molloy–Reed configuration model~\cite{Molloy1995}, in which the degree of each node is drawn according to the target distribution $P(k)$. A number $k_i$ of half-edges (unconnected stubs) is then assigned to each node $i$, and edges are formed by randomly selecting and pairing these stubs.

A power-law distribution $P(k)\sim k^{-\gamma}$ with exponent $2<\gamma<3$ exhibits the so-called natural cutoff, where the average largest value observed in a sample of $N$ events scales as $\av{\kmax} \sim N^{\frac{1}{\gamma-1}}\ll N$~\cite{Boguna2004a}, in which $N$ is the number of nodes. If multiple connections are allowed in the configuration model, the average number of multiedges can be computed as $\left(\av{k^2}/\av{k}-1\right)^2/2$, where $\av{k^n}$ denotes the $n$th moment of the distribution; see Ref.~\cite{Newman2018}, p.~374. If the second moment $\av{k^2}$ is finite, the number of multiedges remains finite, representing a negligible fraction ($\mathcal{O}(N^{-1})$) as the network size increases. On the other hand, for scale-free networks, the second moment and, consequently, the number of multiedges diverge, making their role relevant.

When multiple connection attempts are rejected under a natural cutoff, disassortative degree correlations are induced, whereby high-degree nodes tend to connect preferentially to nodes of smaller degree~\cite{Newman2002}. This effect arises from the fact that pairs of hubs have a non-negligible probability of being selected more than once during the pairing process, leading to repeated rejections of such connection attempts and, consequently, redirecting edges toward lower-degree nodes. Moreover, this limitation becomes increasingly severe for smaller degree exponents, rendering the computational construction of very large networks combinatorially prohibitive, as the remaining unpaired stubs are increasingly likely to belong to hubs that are already connected. On the other hand, if the so-called structural upper cutoff $\kc=\sqrt{\av{k}N}$~\cite{Boguna2004a} is imposed, uncorrelated networks are recovered within the uncorrelated configuration model (UCM)~\cite{Catanzaro2005}.

Using an upper cutoff that scales with system size, one expects relevant quantities to converge to their asymptotic limits as the network size goes to infinity, independently of how the upper cutoff diverges. However, in practice, simulations of dynamical processes on networks are implemented on large but finite systems, where a finite-size scaling analysis becomes imperative~\cite{Lee2013,Boguna2013,Hong2007,Castellano2008,FerrazdeArruda2022,Cota2018}. In particular, epidemic thresholds in scale-free networks, one of the most relevant epidemiological quantities, commonly vanish in the thermodynamic limit~\cite{PastorSatorras2015}, while an effective nonzero threshold can be defined for finite systems~\cite{Ferreira2012,Boguna2013}. However, both disassortative and assortative degree correlations modify the outcomes of epidemic processes on networks~\cite{Moreno2003,Chen2018,Silva2019}. Since simple graphs with scale-free degree distributions and natural cutoff are known to exhibit degree correlations, the structural cutoff has become broadly used for modeling epidemic processes on networks. Moreover, the upper cutoff also changes the finite-size scaling exponents of contact processes on power-law networks~\cite{Boguna2009,Ferreira2011}.

The presence of extremely high degrees in scale-free distributions with a natural cutoff plays an important role even in simple dynamical processes such as the contact process~\cite{Boguna2009}. The absence of correlations can be investigated using annealed networks, where edges are rewired at a high (formally infinite) rate, such that connections exist only in a probabilistic sense~\cite{Boguna2009,Guerra2010}. However, this approach is not suitable for representing static networks, since dynamical correlations between the states of connected nodes are also washed out in the annealed approach. Therefore, an alternative that preserves dynamical correlations while removing structural correlations is to consider multigraphs generated by the configuration model. Note that correlations are absent only at the ensemble level: considering all networks compatible with a given degree sequence, edges are randomly matched and degree and degree correlations vanish upon ensemble averaging~\cite{Dorogovtsev2022}.

The use of scale-free multigraphs to investigate epidemic processes has not been systematically explored, to the best of our knowledge. In the present work, we fill this gap by comparing different epidemic processes on simple graphs and multigraphs with the same degree sequence generated from scale-free degree distributions. We investigate the  properties at the effective epidemic threshold and analyze epidemic localization~\cite{Goltsev2012}, where the network components responsible for triggering and sustaining epidemic activity are quantified using the activity vector~\cite{Silva2021}. We report that, for the susceptible-infected-susceptible (SIS) epidemic model, in which an isolated subgraph composed of a hub and its neighbors (a star subgraph) is able to sustain epidemic activity for a lifetime exponentially long in the hub degree~\cite{Boguna2013}, the presence of multiple edges increases epidemic localization compared with simple graphs. On the other hand, for models in which the star-subgraph lifetime increases no faster than linearly as a function of degree, such as epidemic models with waning immunity whose rate is smaller than the healing rate~\cite{Ferreira2016a}, localization remains almost unaltered, as do the critical epidemic quantities.



\textit{Models and Methods.} We performed simulations on undirected networks with $N$ vertices $i = 1, 2, \ldots, N$ without self-connections. In this work, we focus on an ensemble of networks in which the degree $k_i$ of each vertex $i$ is drawn according to a degree distribution $P(k)$. Networks are constructed following the configuration model~\cite{Molloy1995}: each vertex has $k_i$ unconnected stubs (half-edges); pairs of stubs are randomly selected and connected to form edges; and the process is iterated until all stubs are connected. For simple graphs, if the selected stubs belong to the same vertex or if a connection between the two vertices already exists, the connection is discarded, and two new stubs are randomly selected. In contrast, for multigraphs, multiple connections are allowed. The condition that $\sum_i k_i$ is even is imposed to permit the connection of all edges. In the case of simple graphs, when all remaining stubs belong to vertices that are already connected, the sample is discarded and the iterative process is restarted. We considered degree distributions with minimum degree $k_0 = 3$. For $\gamma > 5/2$, we considered an upper cutoff $k_\text{c} = k_0 N^{\frac{1}{\gamma-1}}$ which scales as the natural cutoff. For $\gamma < 5/2$, we set $k_\text{c} = N^{\frac{1}{\gamma-1}}$ to ensure that the network could be generated with feasible computational resources.


We investigated different spreading dynamics in which each node can be in one of three possible states: susceptible (S), infected (I), or recovered (R). The most general dynamics considered here is the susceptible-infected-recovered-susceptible (SIRS) model~\cite{Keeling_2008}, in which susceptible vertices become infected with rate $\lambda$ per infected neighbor. Infected individuals spontaneously recover with rate $\mu$, becoming temporarily immune. Waning immunity is assumed, such that recovered individuals return to the susceptible state with rate $\alpha$. The susceptible-infected-susceptible (SIS) model is recovered in the limit $\alpha \to \infty$, whereas the susceptible-infected-recovered (SIR) dynamics corresponds to $\alpha = 0$~\cite{Keeling_2008}. We also considered the Contact Process (CP)~\cite{Henkel_2008}, a modified SIS dynamics in which the infection rate associated with each contact is rescaled by the degree $k$ of the infector. For simplicity, we set $\mu = 1$, thereby fixing the time scale.

The transition from the absorbing (frozen) phase to the active (fluctuating) phase occurs at a critical infection rate $\lambda_\text{c}$. All investigated models exhibit an absorbing state~\cite{Henkel_2008}, in which all nodes are susceptible. For finite networks, the system eventually reaches the absorbing state due to stochastic fluctuations, even when the infection rate is above the activation threshold. Therefore, we employ a quasi-stationary (QS) simulation method in which the dynamics is reset to an active configuration whenever an absorbing state is reached~\cite{Sander2016}. We consider the hub reactivation (HR) method, where the most connected vertex (hub) is reinfected whenever the system falls into the absorbing state, except in the localization analysis, in which a randomly selected node is reactivated instead. The reason is that forced activation of the hub artificially induces localization around it, even when the dynamics is deep in the inactive phase. The simulations were implemented using the optimized Gillespie algorithm~\cite{Cota_2017}, as described in the Supplementary Material (SM). The QS quantities were computed after a relaxation time $t_\text{rlx} \in [10^5, 2\times 10^6]$, over an averaging time $t_\text{avg} \in [10^6, 10^7]$, using 10 independent network realizations. 

\begin{figure}[hbt]
	\centering
	\includegraphics[width=0.9\linewidth]{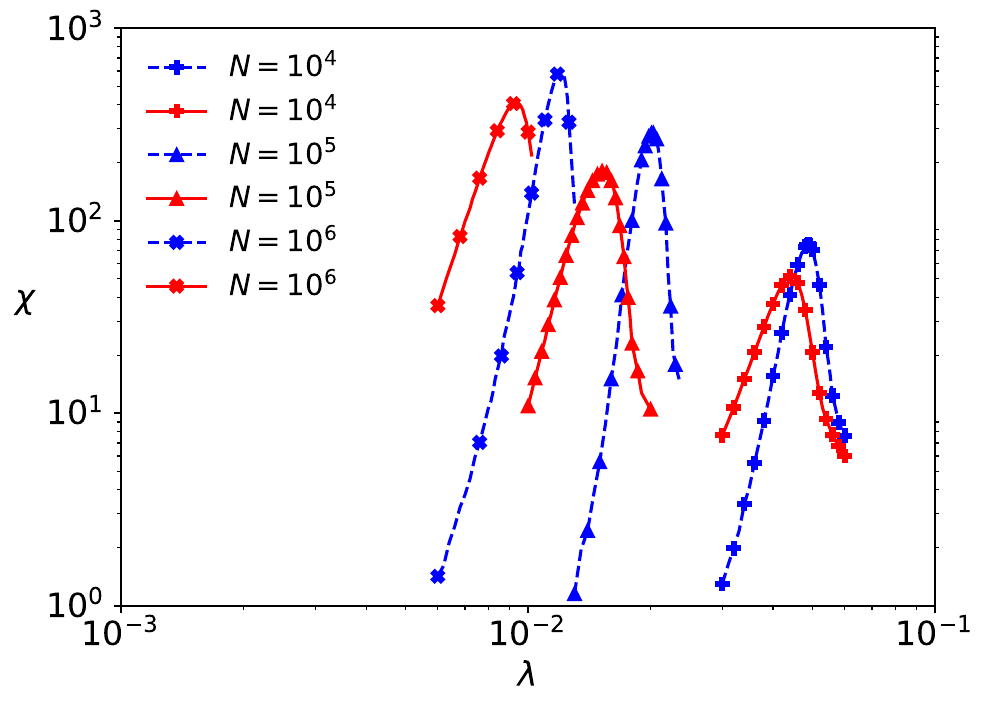}
	\caption{Dynamical susceptibility as function of $\lambda$ for different sizes $N$ for the SIS model for simple (blue, dashed lines) and multigraphs (red, solid lines) with a power-law degree distribution $P(k) \sim k^{-\gamma}$ with $\gamma = 2.6$. }
	\label{fig:suscetibilidadexN}
\end{figure}

The order parameter is the QS density of infected nodes, also called epidemic prevalence. We also investigated the dynamical susceptibility~\cite{Sander2016}, defined as
\begin{equation}
	\chi=N\frac{\av{\rho^2} - \av{\rho}^2}{\av{\rho}},
\end{equation}
which exhibits a peak that diverges with the network size at the transition point and is used as an estimate of the epidemic threshold $\lambda_\text{c}$. Figure~\ref{fig:suscetibilidadexN} compares the dynamical susceptibility for simple graphs and multigraphs of different sizes. Differences between the graph structures can be clearly observed; see Results for further discussion.

We also investigated mean-field approaches for all models in order to assess whether theories originally conceived for simple graphs remain suitable for multigraphs. Since we considered only well-established theories~\cite{Mata2013,Mata2014,Silva2022,Goltsev2012}, the corresponding derivations and results are presented in the SM.

The localization of epidemic processes is investigated using the activity vector $\varrho_i$, defined as the fraction of time that node $i$ remains infected during the QS averaging~\cite{Silva2021}. The normalized activity vector (NAV), given by~\cite{Silva2021}
\begin{equation}
	\phi_i = \frac{\varrho_i}{\sqrt{\sum_{i=1}^N \varrho_i^2}}\,,
\end{equation}
measures the contribution of node $i$ to the order parameter. Localization can be quantified using the inverse participation ratio (IPR)~\cite{Goltsev2012} associated with the corresponding NAV,
\begin{equation}
	Y_4 = \sum_{i=1}^N \left[\phi_i\right]^4.
\end{equation}
If the NAV is localized on a few vertices, $Y_4$ approaches a constant value as $N$ diverges. Conversely, if the NAV is delocalized over the network, $Y_4$ scales as $1/N$. In the intermediate case of subextensive localization, $Y_4$ scales as $N^{-\nu}$ with $0 < \nu < 1$~\cite{Pastor_Satorras_2016}.

\begin{figure}[hbt]
	\centering
	\includegraphics[width=1\linewidth]{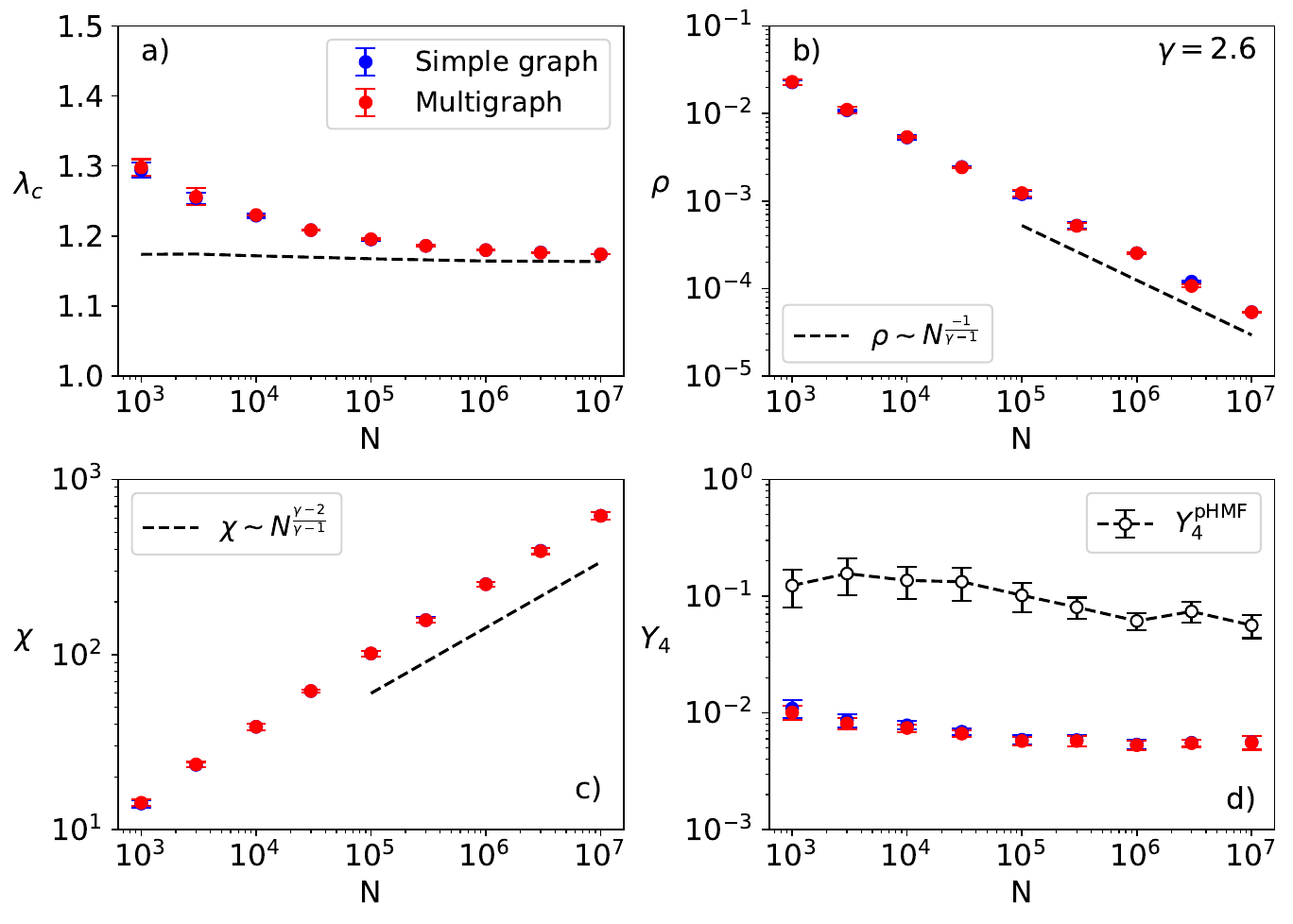}
	\caption{Finite-size scaling of critical quantities obtained from simulations and mean-field theories for the CP on simple graphs and multigraphs with $\gamma=2.6$ and $k_c=3N^{1/(\gamma-1)}$: (a) epidemic threshold, (b) QS density of infected sites, (c) susceptibility, and (d) IPR. Symbols correspond to simulation results, whereas dashed lines indicate the pair HMF prediction of Ref.~\cite{Mata2014} for $\lambda_\text{c}$ and the scaling laws of Ref.~\cite{Silva2021} for $Y_4$.}
	\label{fig:cp_QS}
\end{figure}

\textit{Results}. Figure~\ref{fig:cp_QS} compares the critical properties of the CP dynamics on simple networks and multigraphs for a scale-free degree distribution with exponent $\gamma=2.6$. All investigated quantities are equivalent in both network structures, showing that multigraphs can be employed without significant effects on this model. Contact processes on uncorrelated networks generated with a structural cutoff $k_\text{c}\sim \sqrt{N}$ are very well described by pair heterogeneous mean-field (HMF) theory~\cite{Mata2014}. Simulations with a natural cutoff are also in agreement with pair HMF theory. The density of infected sites scales as $\rho \sim N^{-0.65(2)}$, while the susceptibility scales as $\chi \sim N^{0.39(2)}$. These values agree with the heterogeneous mean-field (HMF) predictions for a natural cutoff, $\rho \sim N^{-\frac{1}{\gamma-1}}$ and $\chi\sim N^{\frac{\gamma-2}{\gamma-1}}$~\cite{Ferreira2011,Mata2014}, indicated in Figs.~\ref{fig:cp_QS}(b) and (c). The activity localization, however, changes drastically when a natural cutoff is adopted: a localized phase characterized by a finite IPR emerges, differing from the subextensive behavior $Y_4\sim N^{-b}$, with $0<b<1$, reported for the CP on networks with a structural cutoff~\cite{Silva2021}. The finite value of $Y_4$ can be analytically explained using the expression derived for the CP within HMF theory~\cite{Silva2021}: $Y_4 ={\langle k^4 \rangle}/{(N \langle k^2 \rangle^2)}$, where $\langle k^i \rangle \sim k_c^{i - \gamma + 1}$ for $i - \gamma + 1 > 0$, and $\langle k^i \rangle \sim \text{const.}$ otherwise. Thus, one can readily verify that $Y_4\sim \text{const.}$ for $k_c \sim N^{1/(\gamma - 1)}$. Indeed, the role of very large hubs (outliers) in altering the critical CP dynamics was previously investigated in annealed networks~\cite{Boguna2009}.

\begin{figure}[hbt]
	\centering
	\includegraphics[width=1\linewidth]{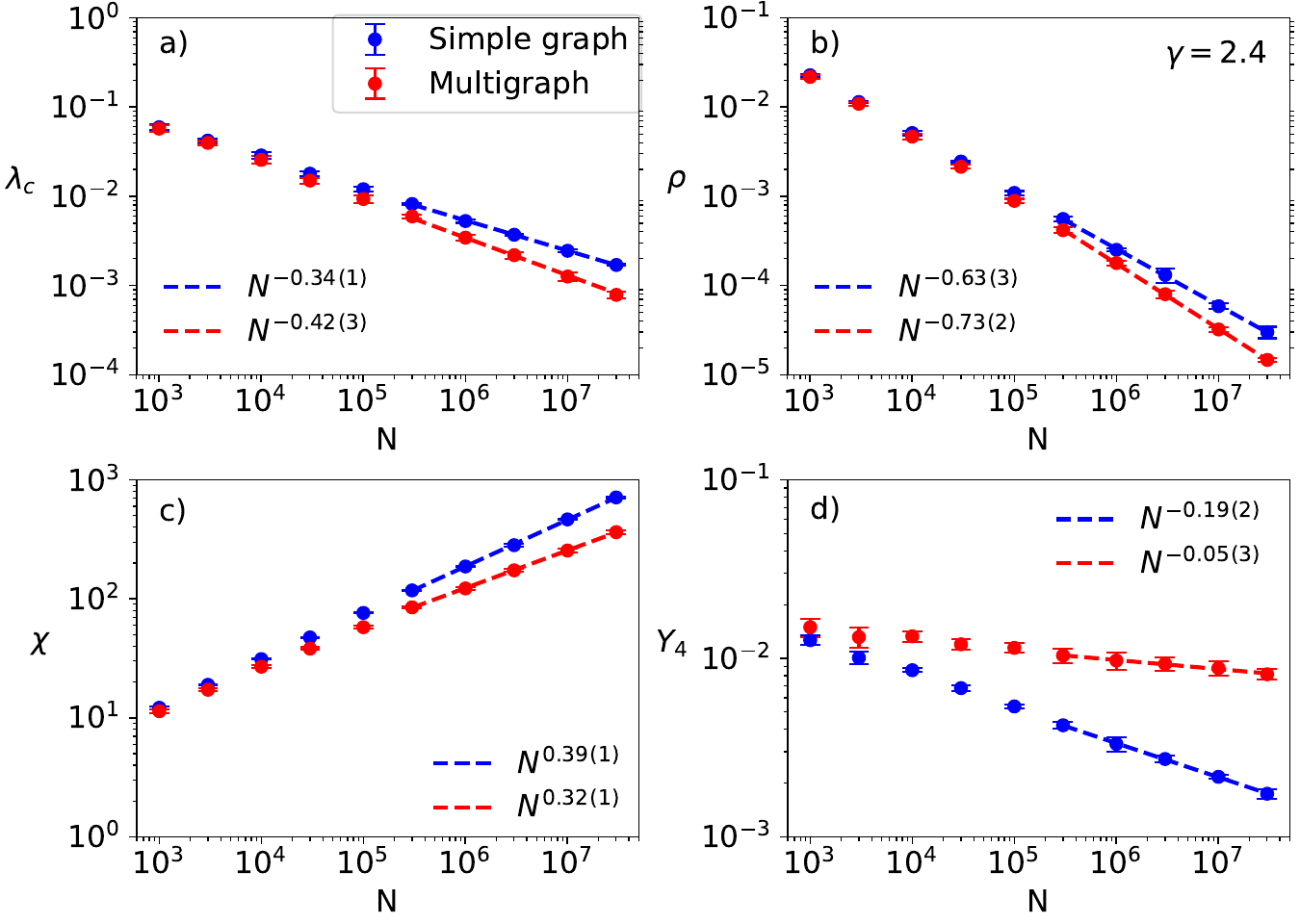}
	\caption{Finite-size scaling for (a) epidemic threshold, (b) epidemic prevalence, (c) susceptibility, and (d) IPR obtained from simulations of the SIS model on simple graphs and multigraphs generated with $\gamma=2.4$ and $k_c=N^{1/(\gamma-1)}$. The QS quantities are analyzed at the epidemic threshold. Dashed lines correspond to power-law fits to the data.}
	\label{fig:sis_g24_QS}
\end{figure}

The finite-size scaling of the SIS model on simple graphs and multigraphs with $\gamma=2.4$, corresponding to the regime in which significant differences in the scaling behavior of the measured quantities emerge, is presented in Fig.~\ref{fig:sis_g24_QS}. Complementary results for $\gamma = 2.6$ and $\gamma = 3$, where these differences are less pronounced, are presented in the SM.

Structural disassortative correlations are strong in simple graphs with $\gamma=2.4$ (see the SM), sufficiently strong to alter the finite-size scaling behavior of the QS quantities compared with that observed in uncorrelated multigraphs, as shown in Figs.~\ref{fig:sis_g24_QS}(b) and (c). The epidemic threshold decays slightly faster in multigraphs, $\lambda_\text{c} \sim N^{-0.42(3)}$, than in simple graphs, for which $\lambda_\text{c} \sim N^{-0.34(1)}$. Localization exhibits a more pronounced change in scaling behavior: for simple graphs, localization is subextensive, with $Y_4 \sim N^{-0.19(2)}$, whereas for multigraphs, the IPR of the NAV scales with network size $N$ with an exponent close to zero; in the latter case, localization on a finite set of nodes cannot be ruled out. Both the epidemic-threshold and IPR scaling behaviors are well described by the pair quenched mean-field (pQMF) theory~\cite{Mata2013}, in which dynamical equations for pairs of connected nodes are used to describe the network dynamics; see the SM.

The observed shifts in the epidemic threshold and changes in localization can be explained by disassortative degree correlations~\cite{Silva2019}: in simple graphs, hubs are surrounded by low-degree nodes, feedback mutual infection of hubs and its neighbors, reducing epidemic localization and consequently promoting more efficient spreading compared with the uncorrelated case. The critical quantities are also slightly affected. The density of infected sites scales as $\rho \sim N^{-0.63(3)}$ for simple graphs and $\rho \sim N^{-0.73(2)}$ for multigraphs, while the susceptibility scales as $\chi \sim N^{0.39(1)}$ for simple graphs and $\chi \sim N^{0.32(1)}$ for multigraphs.

\begin{figure}[hbt]
	\centering
	\includegraphics[width=0.8\linewidth]{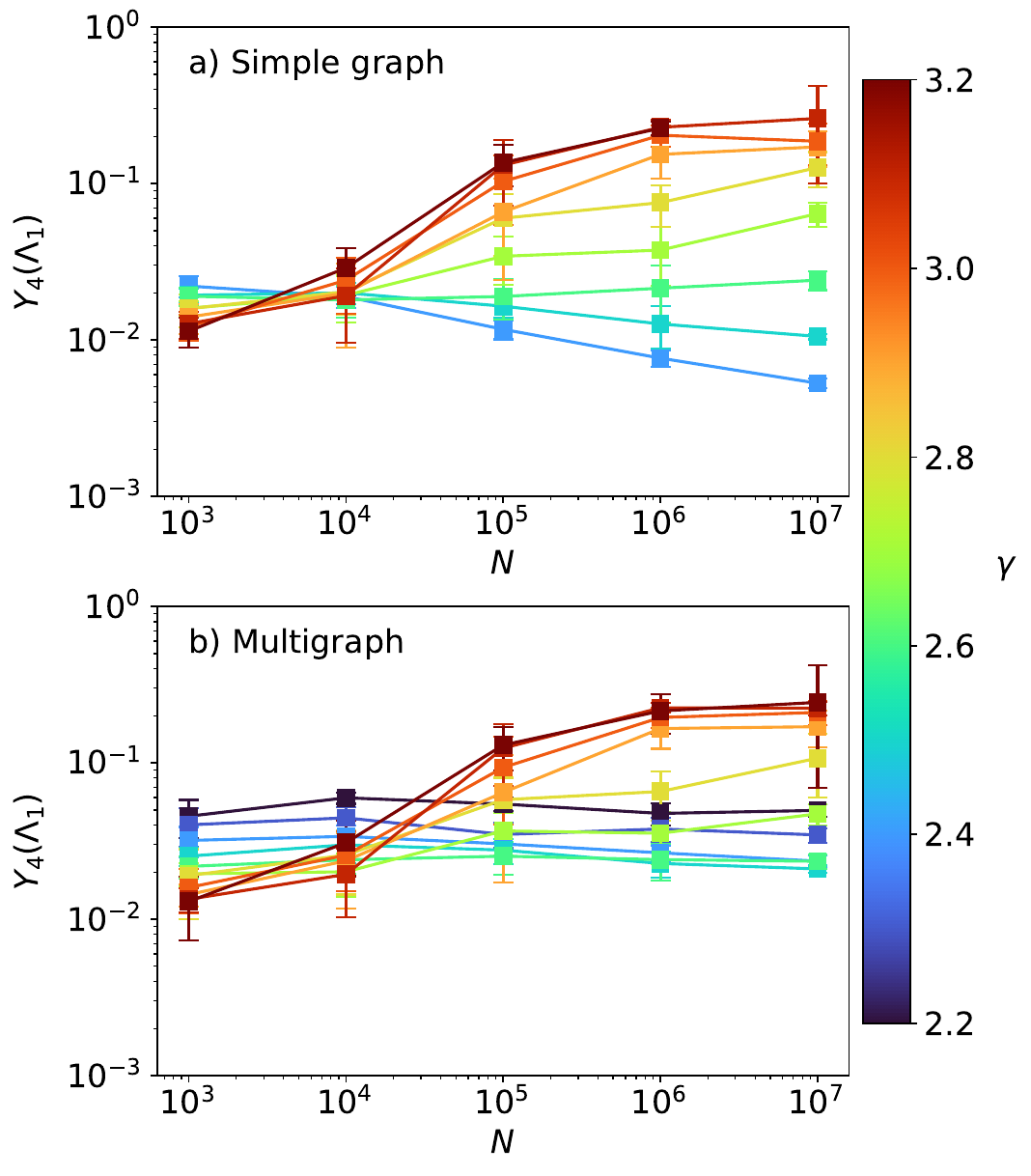}
	\caption{Finite-size scaling of the IPR of the adjacency matrix as a function of network size for different values of $\gamma$, for (a) simple graphs and (b) multigraphs. An upper cutoff $k_\text{c}=N^{1/(\gamma-1)}$ was used for all values of $\gamma$.}
	\label{sis_IPR}
\end{figure}

To gain further insight into how localization depends on the degree exponent, we calculate the IPR of the principal eigenvector (PVE) associated with the largest eigenvalue of the adjacency matrix for simple graphs and multigraphs within the range $2.2 \leq \gamma \leq 3.2$, as shown in Fig.~\ref{sis_IPR}. For simple graphs, only $\gamma \geq 2.4$ was considered due to the computational infeasibility of generating sufficiently large networks with a natural cutoff for $\gamma < 2.4$. Within quenched mean-field (QMF) theory for SIS dynamics, the PVE components are proportional to the probability that a node is infected~\cite{Goltsev2012}; see the SM. Consequently, the PVE corresponds to the normalized activity vector (NAV) in mean-field theory, establishing a direct connection between spectral localization and epidemic activity. For simple graphs, shown in Fig.~\ref{sis_IPR}(a), we recover the behavior previously reported for uncorrelated networks~\cite{Pastor_Satorras_2016}: for $\gamma < 5/2$, the IPR decreases sublinearly with network size and tends to zero in the thermodynamic limit, indicating subextensive localization of the PVE. In contrast, for $\gamma > 5/2$, the IPR converges to a finite value, signaling localization on a finite set of nodes. Multigraphs, however, display markedly different behavior, as presented in Fig.~\ref{sis_IPR}(b). In this case, numerical results consistent with a finite IPR are observed throughout the entire investigated range of $\gamma$, including the regime $\gamma < 5/2$, where simple graphs remain subextensively localized.

This qualitative change can be traced back to the large number of multiple connections established among the major hubs in multigraphs with $\gamma < 5/2$. Since eigenvector centrality favors nodes connected to other highly connected vertices~\cite{Newman2018}, these reinforced hub-to-hub couplings strongly concentrate the PVE around a finite set of dominant nodes. In simple graphs, on the other hand, the prohibition of multiple edges prevents such reinforcement, and localization instead spreads over the densely connected hub core, namely the maximum k-core, which remains subextensive in size~\cite{Pastor_Satorras_2016}. 

\begin{figure}[hbt]
	\centering
	\includegraphics[width=1\linewidth]{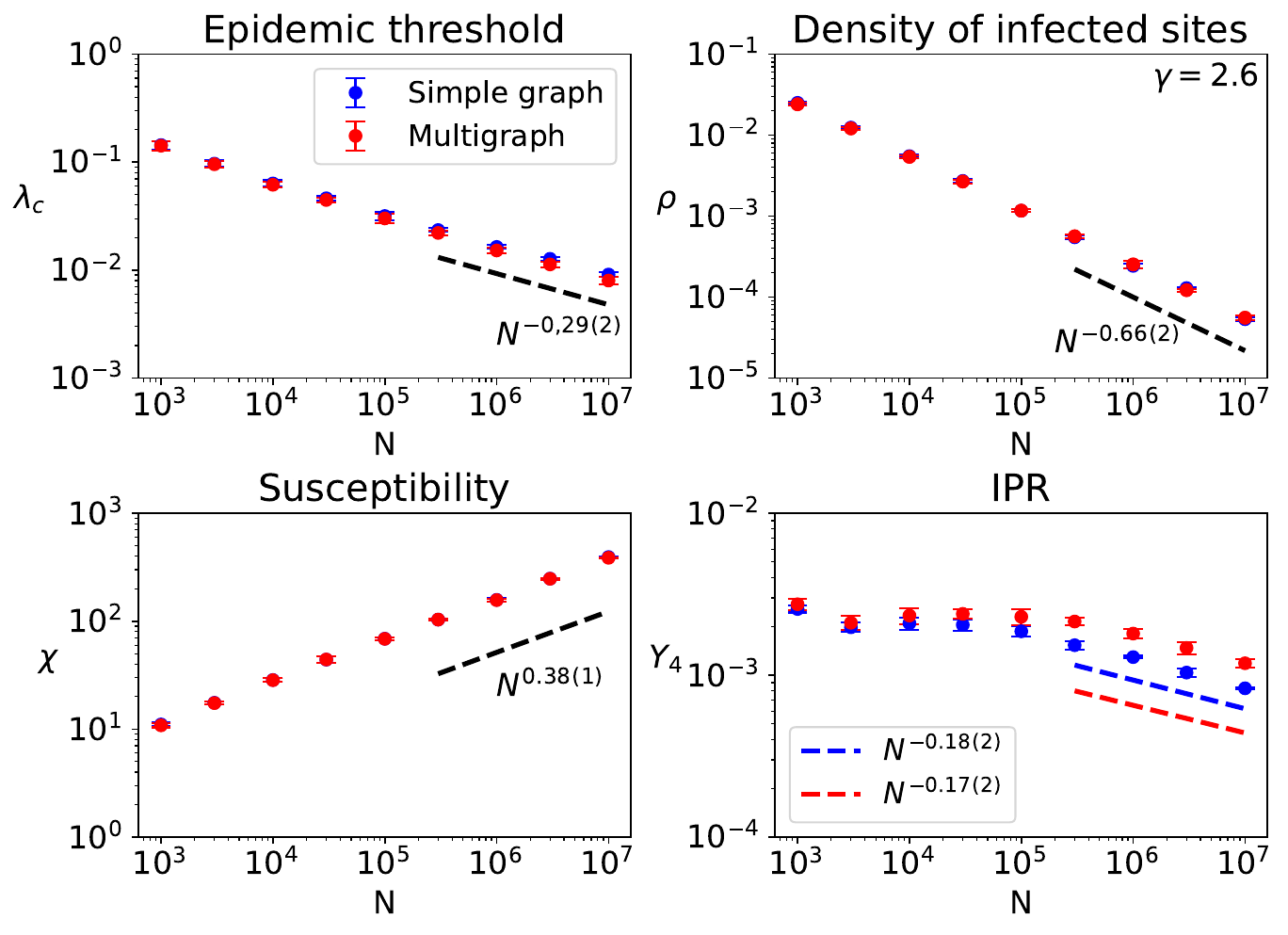}
	\caption{Finite-size scaling of the (a) epidemic threshold, (b) density of infected sites, (c) susceptibility, and (d) IPR obtained from simulations of the SIRS model on simple graphs and multigraphs with degree exponent $\gamma=2.6$ and cutoff $k_c=3N^{1/(\gamma-1)}$. Dashed lines correspond to power-law fits to the data.}
	\label{fig:sirs_g26_QS}
\end{figure}

The effects of waning immunity are investigated using the SIRS model with waning immunity rate $\alpha=0.2$ and degree exponent $\gamma=2.6$, parameters for which differences between the SIRS and SIS models become evident~\cite{Ferreira2016a}. The finite-size scaling for simple graphs and multigraphs is shown in Fig.~\ref{fig:sirs_g26_QS}. Similarly to the contact process case, the finite-size analysis reveals no significant differences in the scaling behavior of either the epidemic threshold or the QS quantities. Regarding localization, the IPR also becomes very similar for multigraphs and simple graphs. However, the natural cutoff introduces an unusual scaling behavior for the IPR, in which a plateau at small sizes, consistent with localization on a finite set of nodes, is followed by an asymptotic decay that scales sublinearly with system size, corresponding to a subextensive component. The pQMF theory for the SIRS model performs worse than for the SIS model in both simple graphs and multigraphs; see the SM. Indeed, it was previously reported that the performance of the pQMF theory for SIRS dynamics deteriorates when the PVE associated with the Jacobian obtained from the stability analysis of the absorbing state becomes more localized~\cite{Silva2022}.

\textit{Conclusions.} In this work, we investigated how multiple connections affect epidemic spreading on scale-free networks by comparing simple graphs and multigraphs generated from the same degree sequence with a natural cutoff. We analyzed several epidemic models exhibiting an active steady state, focusing on finite-size scaling, epidemic thresholds, and localization properties. Our results show that the impact of multiple edges depends fundamentally on the activation mechanism sustaining the epidemic dynamics.

For the contact process, which undergoes collective activation, the critical behavior was essentially unchanged in simple graphs and multigraphs. The epidemic threshold, quasistationary density, susceptibility, and localization properties exhibited the same scaling behavior, indicating that multigraphs can be used without significant effects on the dynamics. For the SIS model, whose activation is driven by hubs, significant differences emerge. Structural disassortative correlations induced by forbidding multiple edges in simple graphs reduce epidemic localization and shift the effective epidemic threshold compared with uncorrelated multigraphs. In particular, multigraphs exhibit stronger localization effects, consistent with epidemic activity concentrated around highly connected hubs. Finally, for the SIRS model with waning immunity, the effects of multiple connections become much weaker, with the SIRS dynamics resembling that of the contact process. The finite-size scaling of all investigated quantities is very similar in simple graphs and multigraphs, consistent with the absence of long-lived hub activation mechanisms.

Overall, our results indicate that multigraphs constitute a useful framework for studying epidemic processes on scale-free networks with a natural cutoff, particularly because they eliminate the structural correlations introduced by forbidding multiple edges and substantially simplify network generation. However, the presence of multiple connections can qualitatively modify localization properties and effective epidemic thresholds whenever the dynamics is dominated by long-lived activity around hubs. These findings clarify the conditions under which multigraphs and simple graphs are equivalent and provide guidance for the appropriate choice of network representations in epidemic modeling.

\begin{acknowledgments}
S.C.F. and F.A.R. acknowledge financial support from the \textit{Conselho Nacional de Desenvolvimento Científico e Tecnológico} (CNPq), Brazil (Grants No. 310984/2023-8, 308162/2023-4, and 407871/2025-0), INCT-NeuroComp (CNPq Grant No. 408389/2024-9), and FAPESP (Grant No. 25/24366-1). S.C.F. and W.C. thank the \textit{Fundação de Amparo à Pesquisa do Estado de Minas Gerais} (FAPEMIG), Brazil (Grants No. APQ-01973-24 and APQ-03079-24). W.C. acknowledges financial support from INCT-DigiSaúde (CNPq Grant No. 408775/2024-6). This study was financed in part by the \textit{Coordenação de Aperfeiçoamento de Pessoal de Nível Superior} (CAPES), Brazil, Finance Code 001.

\end{acknowledgments}

	 %

	 \pagebreak
\clearpage
\onecolumngrid

\begin{center}
	\textbf{\large Supplementary Material: Epidemic spreading on multigraphs}
\end{center}	 

\setcounter{figure}{0}
\renewcommand{\thefigure}{S\arabic{figure}}

\section*{Degree correlations}

\begin{figure}[hbt]
	\centering
	\includegraphics[width=0.7\linewidth]{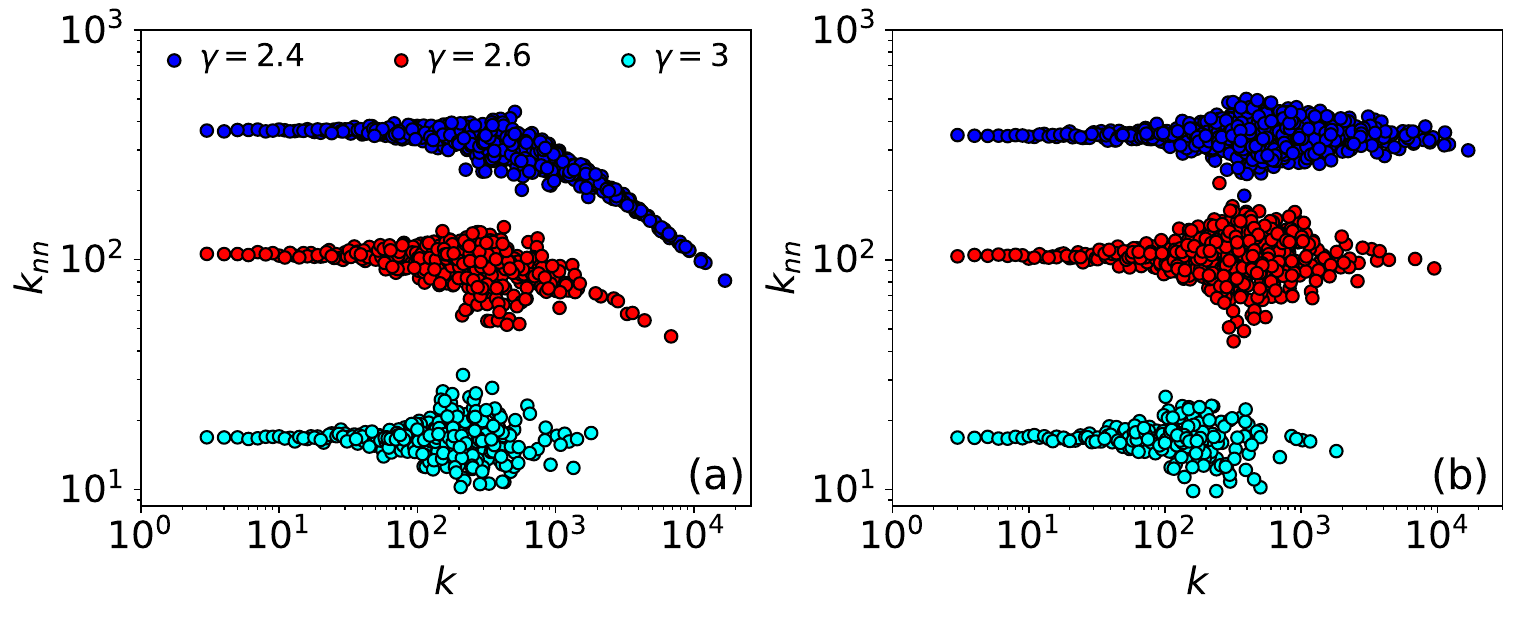}
	\caption{Degree correlations for (a) simple graphs, exhibiting structural disassortativity, and (b) multigraphs, exhibiting neutral correlations. The minimum degree is $k_0=3$, the upper cutoff is $k_\text{c}=N^{1/(\gamma-1)}$ for $\gamma<2.5$, and $k_\text{c}=3N^{1/(\gamma-1)}$, otherwise. Networks have $N = 10^6$ vertices and identical degree distributions for both network types and corresponding parameters. A single network realization is shown.}
	\label{fig:correlation}
\end{figure}

Correlations are characterized using the average nearest-neighbor degree $k_\text{nn}(k)$ as a function of degree $k$~\cite{barabasi2016network}. For uncorrelated networks, $k_\text{nn}(k)$ is independent of $k$ and is given by $\av{k^2}/\av{k}$. For networks generated by the configuration model allowing multiple connections, we observe the expected neutral degree correlations, as shown in Fig.~\ref{fig:correlation}(b). In contrast, simple graphs exhibit structural disassortativity due to the rejection of multiple connections, as shown in Fig.~\ref{fig:correlation}(a), since high-degree nodes are forced to connect to low-degree nodes once connections with other hubs have already been established.

\section*{Stochastic simulations and QS analysis}
	 \label{sec:oga}
The stochastic simulations of epidemic models on networks are implemented using the optimized Gillespie algorithm (OGA)~\cite{Cota_2017}. At each time step, an attempt of recovery, infection, or waning of immunity is performed. The total number of infected individuals $N_{\text{inf}}$, the total number of edges emanating from infected nodes $N_{\text{SI}}$, and the total number of recovered individuals $N_{\text{rec}}$ are determined and continuously updated. The recovery process takes place with probability
	\begin{equation}
		P_{\text{I$\rightarrow$ R} }=\frac{\mu N_\text{inf}}{\mu N_\text{inf} +\lambda N_\text{SI}+\alpha N_{\text{rec}} }\,,
		\label{eq:P_IR}
	\end{equation}
	in which a randomly selected infected node recovers. With probability
	\begin{equation}
		P_{\text{R$\rightarrow$ S}}=\frac{\alpha N_{\text{rec}}}{\mu N_\text{inf} +\lambda N_\text{SI}+\alpha N_{\text{rec}} }\,,
		\label{eq:P_RS}
	\end{equation}
	a randomly selected recovered node returns to the susceptible state. Finally, with probability
		$P_{\text{S$\rightarrow$ I}}=1-P_{\text{I$\rightarrow$ R} }-P_{\text{R$\rightarrow$ S}}$,
an infected node $i$ is selected with probability proportional to its degree. A neighbor $j$ of this node is selected at random, and if it is susceptible, $j$ changes its state to infected; otherwise, the simulation proceeds to the next step. Time is incremented by
\begin{equation}
	\delta t  = \frac{-\ln u}{\mu N_\text{inf}+\lambda N_\text{SI}+\alpha N_{\text{rec}}}\,,
	\label{eq:dt}
\end{equation}
in which $u$ is a pseudorandom number uniformly distributed in the interval $(0,1)$. The SIS dynamics is implemented by setting $N_\text{rec}=0$ in Eqs.~\eqref{eq:P_IR} and \eqref{eq:dt} and assuming that the transition occurs directly from infected to susceptible.

For the contact process, the implementation is simpler: an infected node is selected with equal probability. With probability $p=\lambda/(\mu+\lambda)$, an infection attempt, identical to that in the SIRS dynamics, is performed; otherwise, a healing event, also identical to that in the SIRS dynamics, is performed. The time step is
\begin{equation}
\delta t  = \frac{-\ln u}{ N_\text{inf}(\mu+\lambda)}.
\end{equation}

\section*{Mean-field theories}
\label{sec:mf}
	 
The quenched mean-field theory (QMF)~\cite{Goltsev2012} describes the dynamics of individual nodes and has been applied to the SIS dynamics on power-law networks. The network structure is encoded in the adjacency matrix $\mathbf{A}$, where $A_{ij} = 0,1,2,\ldots$ gives the number of connections between vertices $i$ and $j$. We denote by $k_i$ the degree (number of connections) of vertex $i$, given by $k_i = \sum_{j} A_{ij}$. In simple graphs, $A_{ij}=0$ or 1 is a binary variable. The dynamical equation for the probability that node $i$ is infected is given by
	 \begin{equation}
	 	\frac{d\rho_i}{d t} = -\mu\rho_i+\lambda(1-\rho_i)\sum_{j=1}^{N} A_{ij} \rho_j.
	 	\label{eq:QMFSIS}
	 \end{equation} 
The epidemic threshold is obtained when the largest eigenvalue of the Jacobian
\begin{equation}
L_{ij}=-\mu\delta_{ij}+\lambda A_{ij}
\end{equation}	 
vanishes. This implies that $\lambda_\text{c}=\mu/\Lambda_1$, where $\Lambda_1$ is the largest eigenvalue (LEV) associated with the principal eigenvector (PEV) of the adjacency matrix. The epidemic threshold assumes the same value for the SIRS dynamics, independently of the waning-immunity rate.
	 
A strong approximation in QMF theory is the assumption that the states of neighboring nodes are independent. A significant improvement is achieved by considering dynamical equations at the pair level through the pair quenched mean-field (pQMF) theory~\cite{Mata2013}. The epidemic threshold for the SIRS model can be obtained when the largest eigenvalue of the Jacobian matrix vanishes:
	 \begin{eqnarray}
	 	L_{ij}=-\mu\left[\mu+\lambda k_{j} \Xi(\mu,\lambda,\alpha)\right]\delta_{ij}+\lambda\Upsilon(\mu,\lambda,\alpha) A_{ij},
	 	\label{eq:PQMF_SIRS}
	 \end{eqnarray}
	 where 
	 \begin{eqnarray}
	 	\Upsilon(\mu,\lambda,\alpha)=\frac{2\mu(\mu+\lambda+\alpha)+\lambda\alpha}{2\lambda(\mu+\alpha)+2\mu(\mu+\lambda+\alpha)}
	 	\label{eq:Upsilon}	
	 \end{eqnarray}
	 and
	 \begin{eqnarray}
	 	\Xi(\mu,\lambda,\alpha)=\frac{\lambda(\alpha+2\mu)}{2\lambda(\alpha+\mu)+ 2\mu(\mu+\lambda+\alpha)}.
	 	\label{eq:Xi} 
	 \end{eqnarray}
	 See Ref.~\cite{Silva2022} for the derivation. The SIS limit is recovered for $\alpha\rightarrow\infty$ and agrees with the Jacobian reported in~\cite{Mata2013}.
	 
The epidemic thresholds of the pQMF theories are determined numerically by solving the corresponding transcendental equations using the bisection method to determine the value of $\lambda_\text{c}$ and the power method to evaluate the LEV.  
	 
Slightly above the epidemic threshold, one can show that the local prevalence is given by $\rho_i \sim v_i^{(1)}$, where $v_i^{(1)}$ is the component of the PEV associated with the corresponding Jacobian of the different mean-field theories near the epidemic threshold; see, e.g., Ref.~\cite{Goltsev2012} for SIS QMF theory, Ref.~\cite{Silva2019} for SIS pQMF theory, and Ref.~\cite{Silva2022} for SIRS pQMF theory.	 
	 
\section*{Comparison of pQMF theory with  simulations}

\begin{figure}[hbt]
	 	\centering
	 	\includegraphics[width=0.7\linewidth]{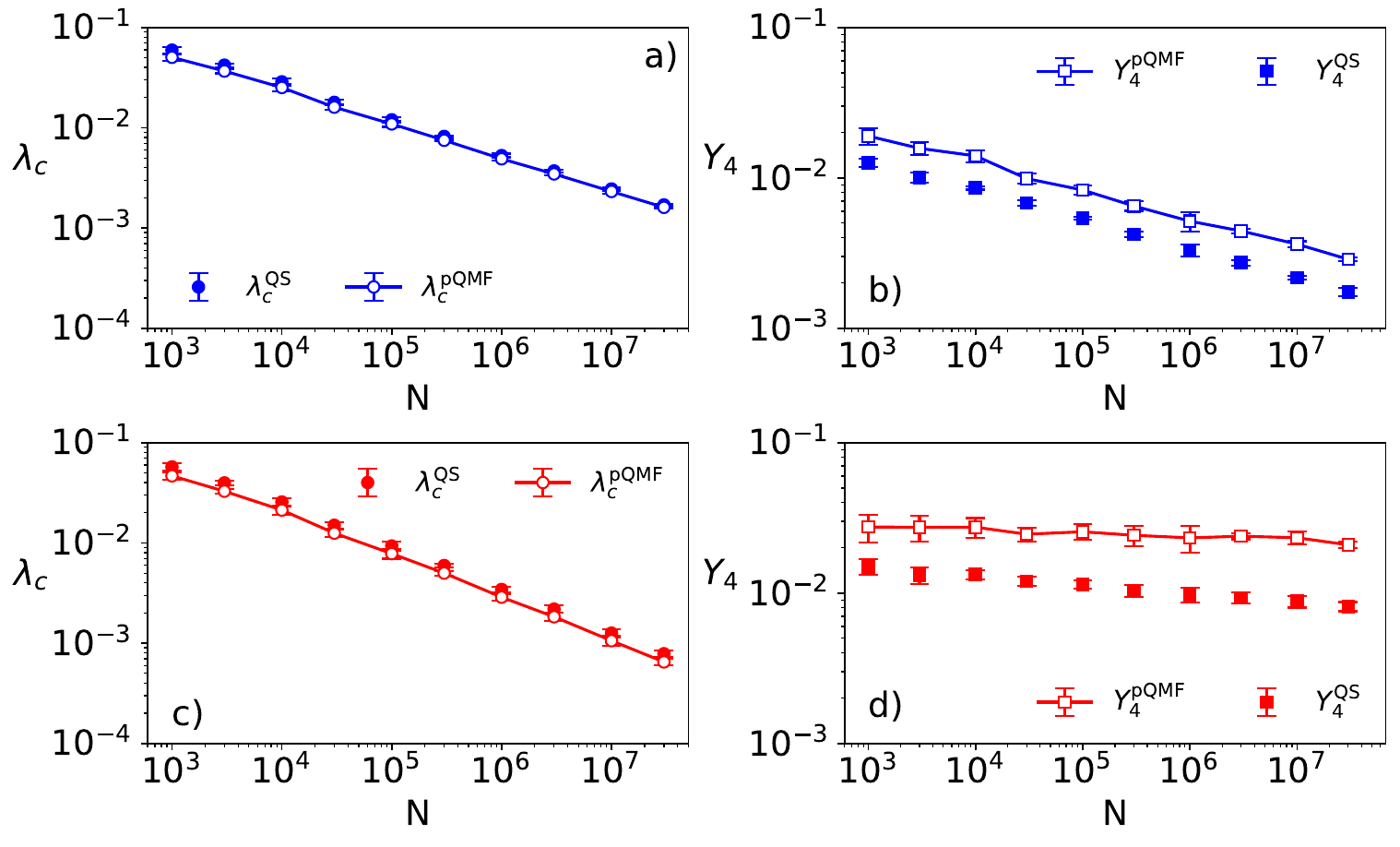}
	 	\caption{Comparison between pQMF theory and QS simulations for the SIS model on (a,b) simple graphs and (c,d) multigraphs with degree exponent $\gamma=2.4$ and cutoff $k_\text{c}=N^{1/(\gamma-1)}$. Finite-size scaling of the (a,b) IPR and (c,d) epidemic threshold is presented.}
	 	\label{fig:sis_g24_MF}
	 \end{figure}

\begin{figure}[hbt]
	 	\centering
	 	\includegraphics[width=0.7\linewidth]{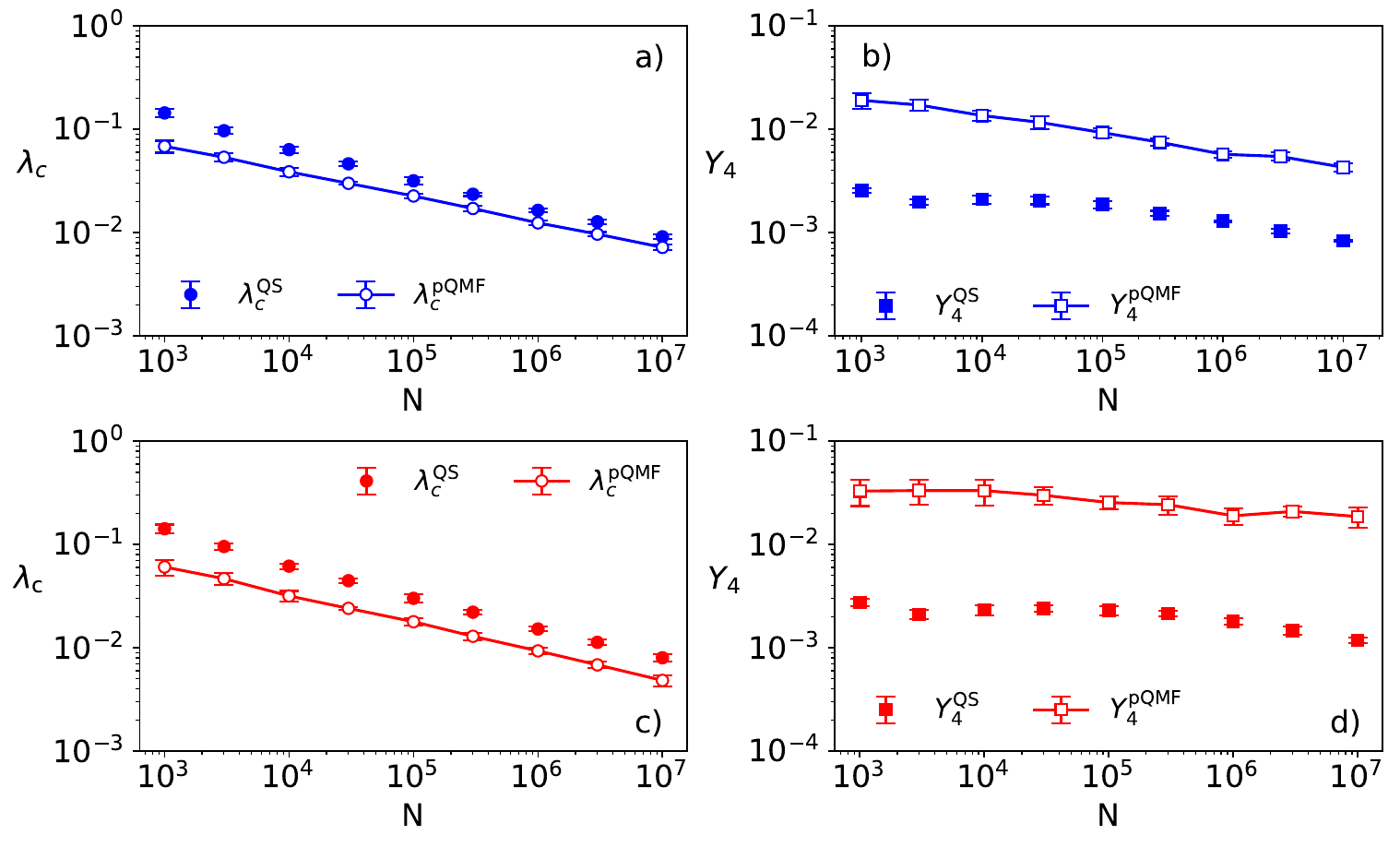}
	 	\caption{Comparison between pQMF theory and QS simulations for the SIRS model on (a,b) simple graphs and (c,d) multigraphs with degree exponent $\gamma=2.6$ and cutoff $k_\text{c}=3N^{1/(\gamma-1)}$. Finite-size scaling of the (a,b) IPR and (c,d) epidemic threshold is presented.}
	 	\label{fig:sirs_g26_MF}
	 \end{figure}

Figures~\ref{fig:sis_g24_MF} and \ref{fig:sirs_g26_MF} present comparisons between pQMF theory and QS simulations for the SIS and SIRS models, respectively. 
	 
	 \section*{Complementary results}
	 
\begin{figure}[hbt]
	 	\centering
	 	\includegraphics[width=0.7\linewidth]{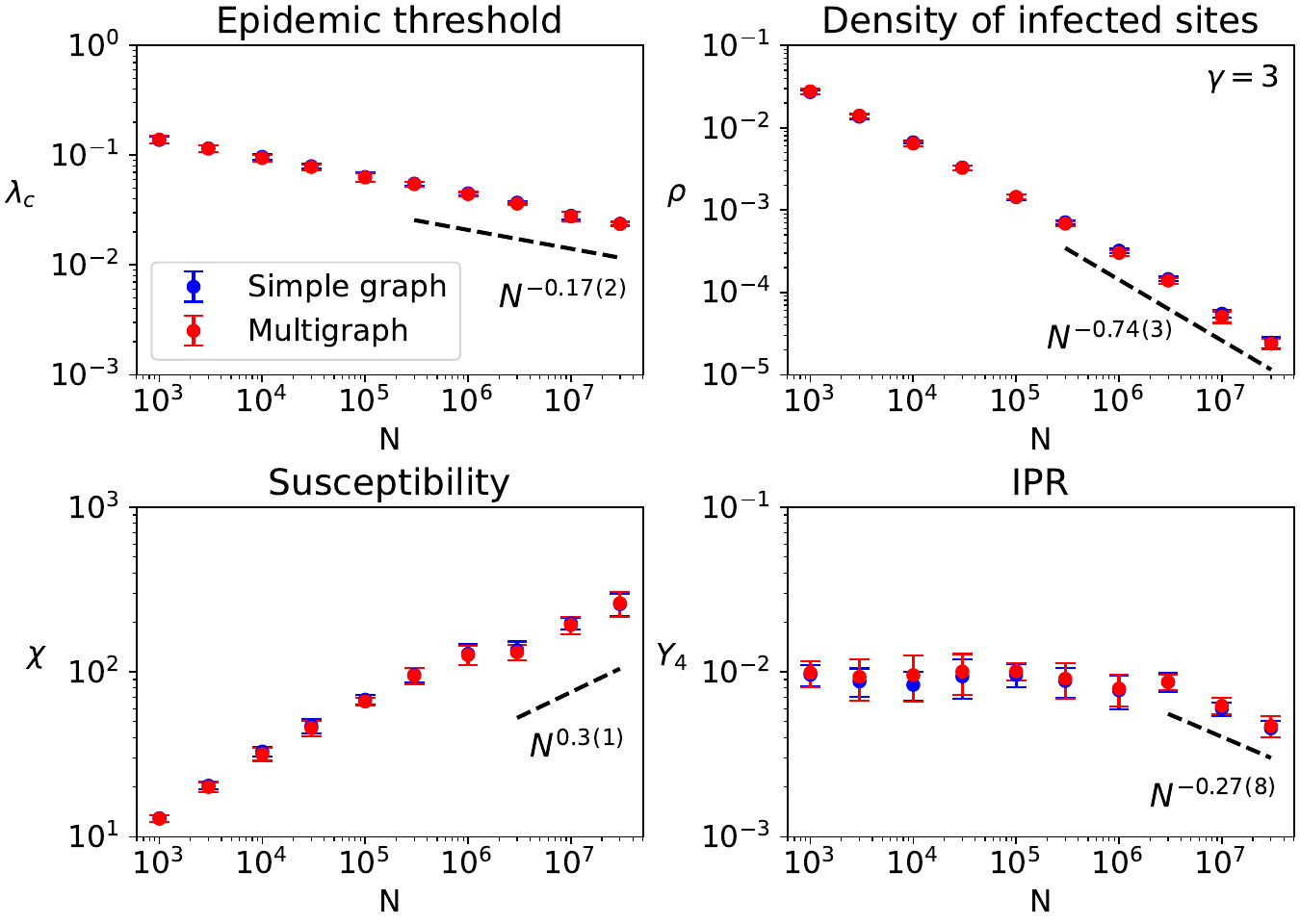}
	 	\caption{Finite-size scaling of QS quantities for SIS dynamics on simple graphs and multigraphs with $\gamma=3$ and $k_c=3N^{1/(\gamma-1)}$: (a) epidemic threshold, (b) density of infected sites, (c) susceptibility, and (d) IPR. Dashed lines correspond to power-law fits to the data.}
	 	\label{fig:sis_g3_QS}
	 \end{figure}

\begin{figure}[hbt]
	 	\centering
	 	\includegraphics[width=0.7\linewidth]{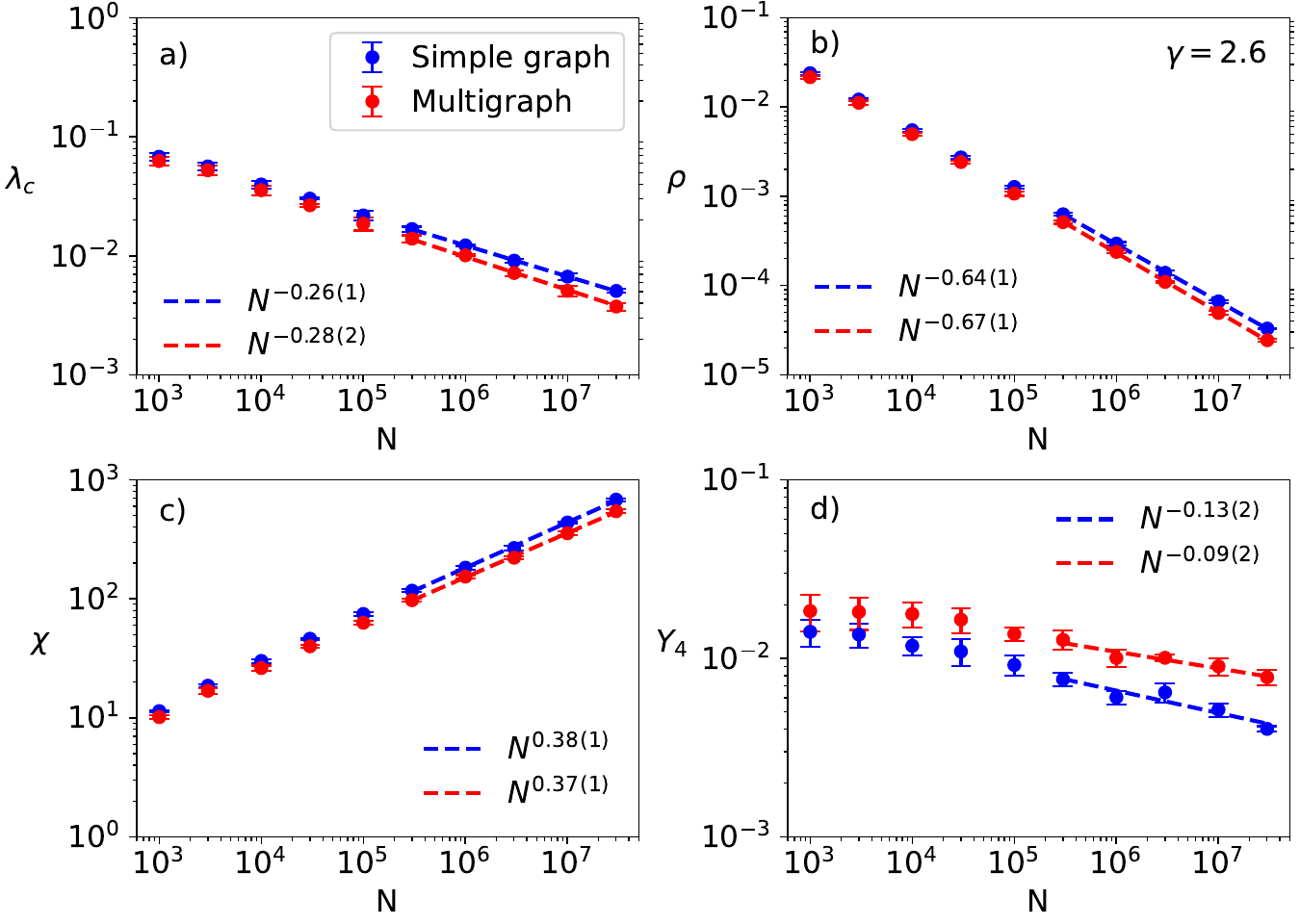}
	 	\caption{Finite-size scaling of QS quantities for SIS dynamics on simple graphs and multigraphs with $\gamma=2.6$ and $k_c=3N^{1/(\gamma-1)}$: (a) epidemic threshold, (b) density of infected sites, (c) susceptibility, and (d) IPR. Dashed lines correspond to power-law fits to the data.}
	 	\label{fig:sis_g26_QS}
	 \end{figure}

Figures~\ref{fig:sis_g3_QS} and \ref{fig:sis_g26_QS} compare the SIS dynamics on simple graphs and multigraphs for degree exponents $\gamma=3$ and $\gamma=2.6$, respectively. For $\gamma=3$, both network structures yield equivalent results, since multiple edges are negligible. For $\gamma=2.6$, small differences can be resolved, although they are weaker than those observed for $\gamma=2.4$ in the main text.
	 
\end{document}